\documentclass[a4paper,fleqn,usenatbib]{mnras}

\usepackage[T1]{fontenc}
\usepackage{ae,aecompl}

\usepackage{graphicx}	
\usepackage{amsmath}	
\usepackage{amssymb}	
\usepackage{xcolor}

\title[Preprocessing, mass loss and mass segregation]{Preprocessing, mass loss and mass segregation of galaxies in DM simulations}

\author[G.D. Joshi, J. Wadsley and L. C. Parker]{Gandhali D. Joshi\thanks{Email: joshigd@mcmaster.ca}, James Wadsley, Laura C. Parker \\
Department of Physics and Astronomy, McMaster University, Hamilton, ON L8S 4M1, Canada}

\date{Accepted XXX. Received YYY; in original form ZZZ}

\pubyear{2016}

\begin{document}
\label{firstpage}
\pagerange{\pageref{firstpage}--\pageref{lastpage}}
\maketitle

\begin{abstract}
{We investigate the mass loss of galaxies in groups and clusters with high-resolution DM simulations. We detect weak mass segregation in the inner regions of group/cluster haloes, consistent with observational findings. This applies to samples of galaxy analogues selected using either their present-day mass or past maximum (peak) mass. We find a strong radial trend in the fractional mass \emph{lost} by the galaxies since peak, independent of their mass. This suggests that segregation is due to massive galaxies having formed closer to the halo centres and not the preferential destruction of smaller galaxies near halo centres. We divide our sample into galaxies that were accreted as a group vs. as a single, distinct halo. We find strong evidence for preprocessing -- the grouped galaxies lose $\sim 35-45\%$ of their peak mass before being accreted onto their final host haloes, compared to single galaxies which lose $\sim12\%$. After accretion, however, the single galaxies lose \emph{more} mass compared to the grouped ones. These results are consistent with a scenario in which grouped galaxies are preprocessed in smaller haloes while single galaxies `catch up' in terms of total mass loss once they are accreted onto the final host halo. The fractional mass loss is mostly independent of the galaxy mass and host mass, and increases with amount of time spent in a dense environment.}
\end{abstract}




\section{Introduction}	\label{sec:intro}
Galaxies have been shown to be influenced by their environment in different ways -- on average, dense environments host larger fractions of galaxies that are red, elliptical and quenched, compared to isolated galaxy populations \citep[e.g.][]{Oemler74,Dressler80,Balogh04,Hogg04,Kauffmann04,Blanton05}. 
Not only are galaxy populations different in groups/clusters compared to the field, galaxy properties can also depend on their group/cluster-centric radius. Such a radial segregation has been seen in properties such as morphology and luminosity \citep[e.g.][]{Girardi03}, colour \citep[e.g.][]{Blanton07}, quenched fractions \citep[e.g.][]{Wetzel12} and star formation rates \citep[e.g.][]{Balogh00}. The observed segregation in these properties can point towards various pathways of galaxy evolution. Furthermore, the question of where galaxies begin to be affected by their environment is yet to be answered and may be a major factor in establishing such trends.

Several different mechanisms have been put forth to explain galaxy properties in groups and clusters; processes such as mergers \citep{Makino97,Angulo09}, ram pressure stripping by the intra-group/cluster medium \citep{Gunn72,Abadi99}, harassment by other galaxies \citep{Moore96,Moore98} and starvation \citep{Larson80,Balogh00,Kawata08} can each affect galaxies differently. Each of these processes is efficient on different timescales and at different masses and halo-centric radii. While the properties mentioned above all involve the baryonic content of the galaxy, mass segregation could arise purely through the gravitational interactions of its dark-matter halo with other haloes. Understanding mass segregation can therefore point to which of the above processes are dominant in group and cluster environments. 

Several observational and simulation studies have examined mass segregation and have found conflicting evidence. On the observational side, \citet{Lares04} found significant segregation trends in the velocity functions of galaxies in different luminosity (and therefore stellar mass) ranges using data on group galaxies from the 2dFGRS \citep{Colless01}. \citet{vanDenBosch08} found segregation in stellar mass using data from the SDSS-DR7 \citep{Abazajian09} and concluded that the trends were a result of mass segregation and correlations between stellar mass and colour/concentration. More recently, \citet{Roberts15} found weak stellar mass segregation in galaxy groups in the SDSS-DR7. They also found that the segregation trends were stronger when low mass galaxies were included and that the trends were weaker in higher mass groups/clusters. In contrast to these results, \citet{vonDerLinden10} found no evidence of mass segregation in cluster galaxies in the SDSS, while \citet{Ziparo13} only found mild stellar mass segregation and only at low-redshift using X-ray selected groups from the COSMOS, GOODS and ECDFS fields. Most recently, \citet{Kafle16} also found no evidence of mass segregation using data from the GAMA survey \citep{Driver11}, the GALFORM semi-analytic model \citep{Gonzalez14}, and the EAGLE simulation \citep{Schaye15} in a wide range of halo masses.  Using simulations and semi-analytic modelling, \citet{Contini15} found significant mass segregation trends; they found that average mass decreased with halo-centric radius out to $R_{\text{vir}}$, then increased with radius out to $2\,R_{\text{vir}}$. \citet{vanDenBosch16} conducted an extensive study of segregation in various properties of dark matter haloes in the Bolshoi and Chinchilla simulations. While they only found a mild correlation in the present-day masses of the haloes as a function of their halo-centric radius, they did find much stronger correlations when considering the mass at accretion or the amount of mass lost after accretion.

What drives mass segregation, if it does exist, is also not completely understood and several effects could be at work. Dynamical friction \citep{Chandrasekhar43}, where the drag force from surrounding matter preferentially drives more massive haloes towards the centres of groups/clusters, may be one of the more important ones, as has been predicted by several studies \citep[e.g. see][]{Ostriker75,Tremaine75,White77}. Galaxies that are accreted earlier will also be located at smaller radii since virial radii at high redshifts were smaller; if these galaxies are also preferentially higher in mass, this can lead to mass segregation. Additionally, each of these effects have to contend with galaxies losing or gaining mass due to tidal stripping and mergers. Observing mass \emph{loss} in particular would imply that tidal interactions, both with the encompassing host halo as well as the haloes of other galaxies in the group/cluster, are important in these environments.

These issues regarding mass loss of galaxy haloes also lead to the phenomenon of preprocessing. While all of the processes mentioned above may occur in the current host haloes, there is mounting evidence that some of these galaxies were accreted as part of smaller groups and that any response they have had to their environment began in these smaller groups. \citet{McGee09} studied cluster assembly histories using the Millennium simulation and semi-analytic modelling and concluded that a large fraction of galaxies were accreted onto these clusters as part of smaller groups and were therefore potentially preprocessed. They did find, however, that the degree of preprocessing was dependent on the stellar mass of the infalling galaxies whereby more massive galaxies were more likely to be preprocessed. \citet{Bahe13} studied radial trends in cold and hot gas mass and star formation rates (SFRs) in clusters in the GIMIC suite of simulations. They found that $\sim50\%$ of galaxies in massive clusters were accreted as part of smaller groups, decreasing to $<10\%$ in low-mass groups, and that when these preprocessed galaxies were excluded, most of the radial trends they had observed were significantly weakened. \citet{Hou14} used group and cluster galaxies in the SDSS-DR7 and found that galaxies in `subhaloes' (i.e. galaxy clumps within the group or cluster) showed enhanced quenched fractions beyond $\sim1.5-2\,r_{200}$. They also concluded that preprocessing was important in massive clusters, but less so in smaller groups. \citet{Gabor15} studied clusters in a hydrodynamical simulation and found $\sim1/3$ of the cluster galaxies had been quenched in groups of mass $>10^{12}M_{\sun}$. 

It is important to establish whether preprocessing is indeed an important factor in the observed galaxy properties in groups and clusters. The environmental processes discussed above can alter galaxy properties; however, different mechanisms are dominant in different environments and determining where a galaxy begins transforming can shed light on what process(es) will drive its transformation. Preprocessing may account for a majority of the differences we observe between field and cluster galaxies, especially in the case of massive clusters that are expected to have accreted several smaller groups. Being able to separate such galaxies that have been preprocessed in groups from those that are truly influenced by the cluster will help clarify how clusters influence their member galaxies. Preprocessing also plays an important role in determining whether we see segregation in certain environments -- while segregation may occur in smaller groups, as these groups are accreted onto larger clusters, they may destroy any trends formed within the cluster itself.

In this study, we use dark matter simulations to study both mass segregation and preprocessing in terms of mass loss and gain, and how these processes depend on various properties of the galaxy haloes. The paper is organized as follows: in Section \ref{sec:methods} we describe the simulations and the data used for this study. Section \ref{sec:massSeg} examines mass segregation in terms of present-day mass as well as peak mass and motivates the use of peak mass for further analysis. In Section \ref{sec:preproc}, we study the role of preprocessing and its dependence on various properties of the galaxy haloes. Finally, we discuss our results in Section \ref{sec:disc} and summarize our findings in Section \ref{sec:summ}.

\section{Methods}	\label{sec:methods}

\subsection{Simulation}	\label{sec:code}
The data for this study come from a cosmological dark-matter simulation of a $\left(100\,\text{Mpc}\right)^3$ comoving volume run using the Tree-SPH code \textsc{Changa} in N-body mode \citep{Jetley08,Jetley10,Menon14}. Initial conditions were generated using the code \textsc{Music} \citep{Hahn13} assuming a flat standard  $\Lambda$CDM cosmology with $\Omega_{\Lambda} = 0.6914$, $\Omega_{m} = 0.3086$, $h = 0.6777$, $n_{s} = 0.9611$ and $\sigma_{8} = 0.8288$ \citep{Planck14}. The simulation contains $1024^3$ particles resulting in a particle mass of $3.7\times10^{7}\,M_{\sun}$ and a gravitational softening length of $1.25\,\text{kpc}$. The softening length is physical for $z<8$ and comoving at higher redshifts. The simulation was run from $z=100$ to $z=0$ in $1000$ timesteps with every $5$\textsuperscript{th} timestep saved, thus producing $200$ snapshots equally spaced in time with each pair of consecutive snapshots separated by $\sim68.9\,\text{Myr}$. Bound haloes were identified in each snapshot with the phase-space Friends-Of-Friends (FOF) algorithm \textsc{Rockstar} \citep{Behroozi13}, and used to generate particle-based merger trees which were further refined using the code \textsc{Consistent Trees} \citep{Behroozi13b}. All halo properties are defined within a virial radius inside which the average density is $\Delta_{\text{c}}$ times the critical density of the Universe. The overdensity factor $\Delta_{\text{c}}$ is obtained for a flat $\Lambda$CDM Universe following \citet{Bryan98} as 
\begin{equation}
\Delta_{\text{c}}(z) = 18\upi^2+82x-39x^2
\end{equation}
where
\begin{equation}
x = \frac{\Omega_{\text{m,0}}(1+z)^3}{\Omega_{\text{m,0}}(1+z)^3+\Omega_{\Lambda}}-1
\end{equation}
For the cosmological parameters used in this study, this gives $\Delta_{\text{c}}=102$ at $z=0$.

\subsection{Galaxy analogues}
We begin by selecting a population of galaxy analogues -- haloes that would host observed galaxies regardless of their position in the extensive subhalo hierarchy produced by the halo finder. In previous work \citep{Joshi16}, we explored the differences between two halo finders in detecting substructure and the importance of taking into account the entire subhalo hierarchy in order to select a sample of haloes that could potentially host galaxies. We therefore defined a simple set of selection criteria that account for the portion of haloes missed due to their being embedded deep in the subhalo hierarchy. Briefly, starting at the distinct halo at the top of the subhalo hierarchy, we sort each candidate halo through the following criteria:
\begin{enumerate}
	\item If $M_{\text{halo}}<10^{10}\,M_{\sun}$, the halo and its branches in the hierarchy are eliminated.
	\item If $M_{\text{halo}}>10^{12.5}\,M_{\sun}$, the halo itself is eliminated, but each of its subhaloes is put through criteria (i)-(iv).
	\item If $10^{10}<M_{\text{halo}}<10^{12.5}\,M_{\sun}$ and the halo has no subhaloes with $M_{halo}>10^{10}\,M_{\sun}$, the halo is accepted as an analogue while its subsequent branches in the hierarchy are eliminated.
	\item If $10^{10}<M_{\text{halo}}<10^{12.5}\,M_{\sun}$ and the halo has at least one subhalo with $M_{halo}>10^{10}\,M_{\sun}$, then we consider $M_{\text{rem}} = \left(M_{\text{halo}}-\sum M_{\text{subhalo}}\right)$; if $10^{10}<M_{\text{rem}}<10^{12.5}\,M_{\sun}$, then the halo is accepted as a galaxy analogue while each of its subhaloes is also put through criteria (i)-(iv).
\end{enumerate}
The mass limits used for these criteria were chosen to correspond to the stellar masses of galaxies in typical redshift surveys, as we show in  Section \ref{sec:obs}. We generate two sets of galaxy analogues, one where the mass under consideration is the $z=0$ virial mass and one where it is the peak mass of the halo. A total of $43530$ galaxy analogues were selected for the first set, and $80425$ for the second. For the remainder of this paper, we refer to the galaxy analogues as just `galaxies', although we emphasize that this nomenclature is adopted for simplicity -- we are only analysing dark matter haloes in this paper, and therefore make no statements regarding the fate of their stellar or gas content. Note that the large difference we see in the sizes of the two samples is because the $M_{\text{peak}}$-selected sample contains galaxies that had relatively low peak masses that have subsequently lost mass and by present-day, fall below the lower mass threshold to be included in the $M_{\text{z=0}}$-selected sample; this affects $\sim46.5\%$ of the $M_{\text{peak}}$-selected sample. Conversely, high mass galaxies in the $M_{\text{z=0}}$-selected sample could also have peak masses beyond the upper mass limit to be included in the $M_{\text{peak}}$-selected sample; however, the number of such cases is negligibly low compared to the first case, $\sim0.3\%$ of the $M_{\text{z=0}}$-selected sample.

\section{Mass segregation} \label{sec:massSeg}

\begin{figure*}
\includegraphics[width=\textwidth]{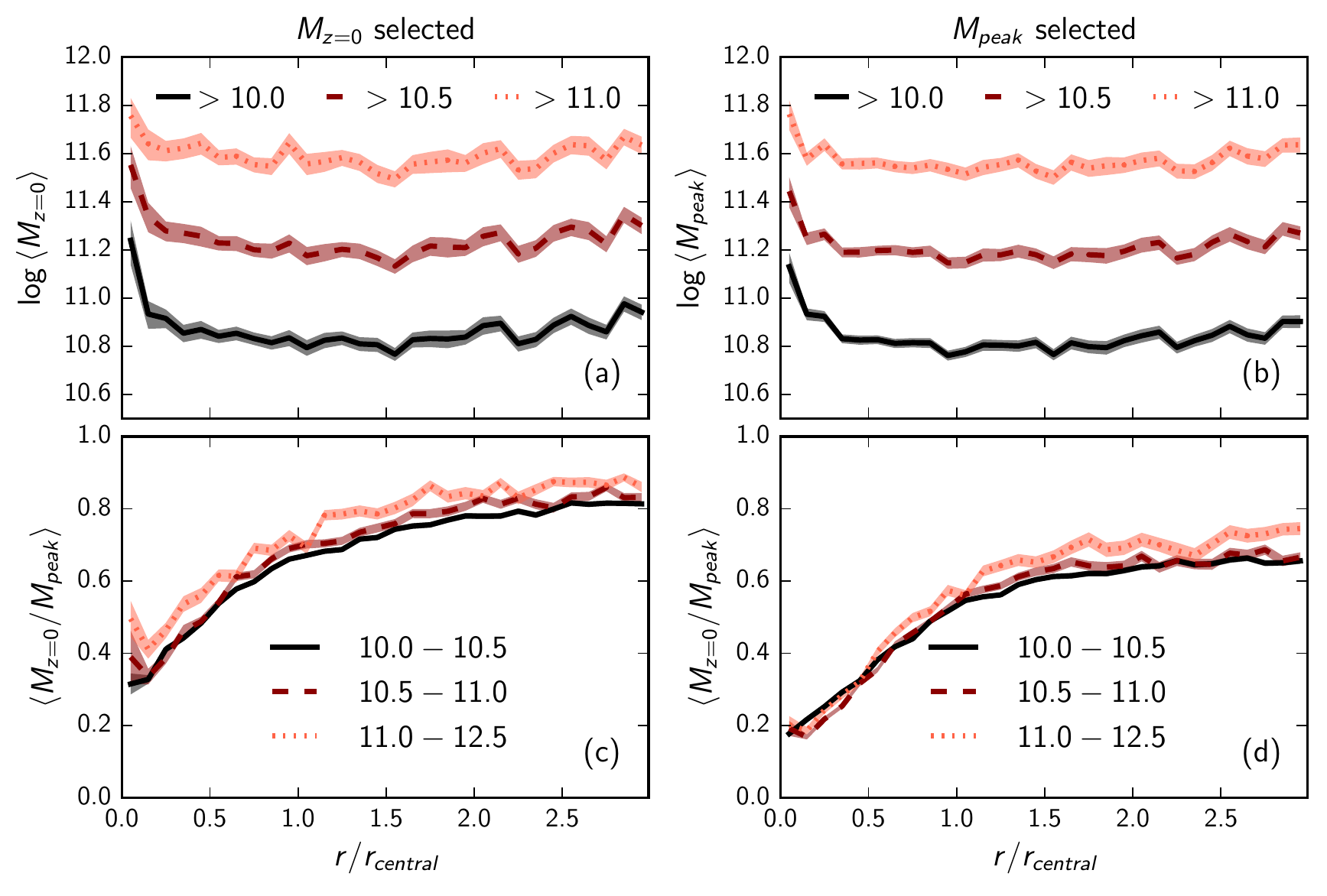}
\caption{Segregation in DM-halo mass of galaxies. \emph{(a)} \& \emph{(b)} show the average mass of the galaxies as a function of present-day halo-centric radius with mass defined as present-day mass, $M_{\text{z=0}}$, in \emph{(a)} and peak mass, $M_{\text{peak}}$, in \emph{(b)}. The different colours (and linestyles) represent bins of $\log{M_{\text{z=0}}}$ in the \emph{(a)} and $\log{M_{\text{peak}}}$ in \emph{(b)}; the upper limit is always set to $12.5$ and the lower limit is altered to examine the effect of excluding low mass galaxies. \emph{(c)} \& \emph{(d)} show the average fractional mass retained from peak to present day, $M_{\text{z=0}}/M_{\text{peak}}$, as a function of halo-centric radius. The colours represent bins of $\log{M_{\text{z=0}}}$ in \emph{(c)} and $\log{M_{\text{peak}}}$ in \emph{(d)}.  The errors shown are standard errors on the mean.} \label{fig:massSegMbins}
\end{figure*}

\begin{figure*}
\centering
\includegraphics[width=\linewidth]{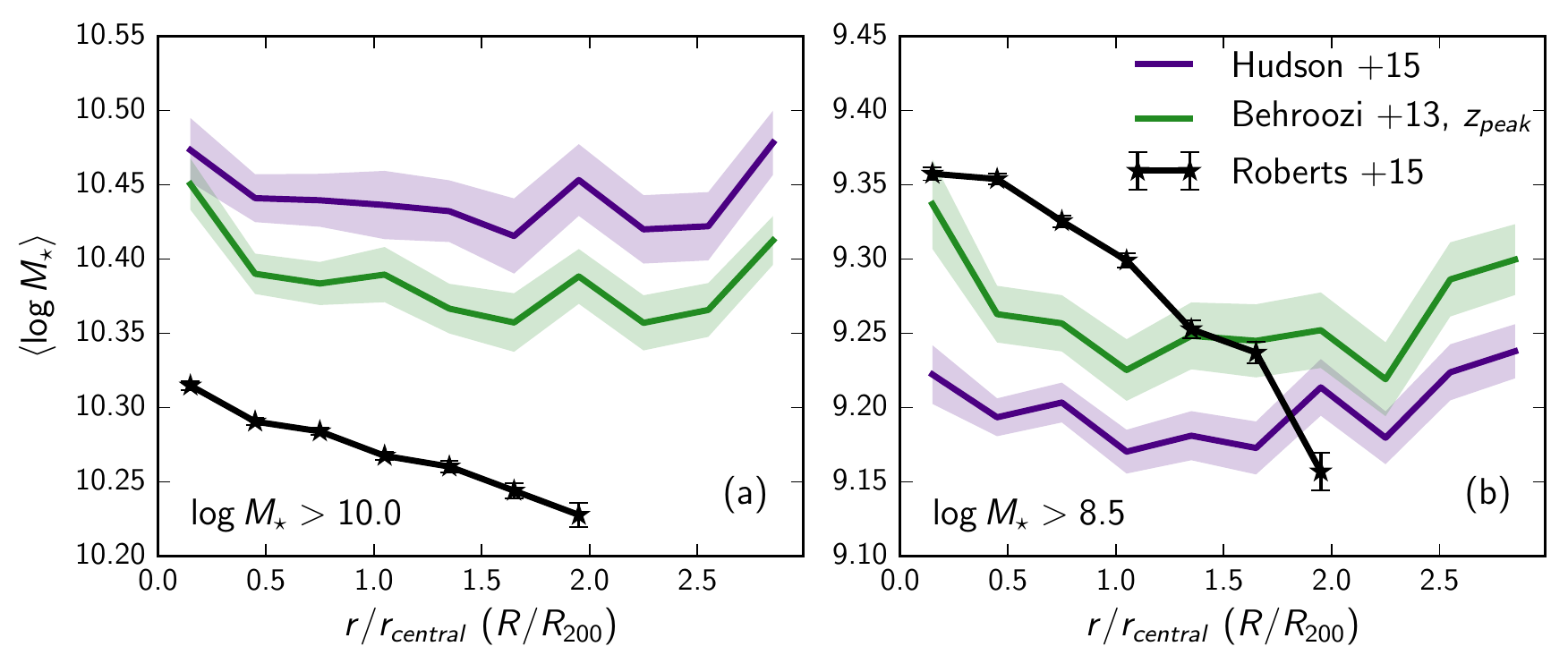}
\caption{Segregation in stellar mass. Average stellar mass as a function of halo-centric radius using two different stellar-mass-to-halo-mass prescriptions: the purple line uses the `no-evolution' fit from \citet{Hudson15}; the green line uses the prescription from \citet{Behroozi13c} defined at $z_\text{{peak}}$ for each galaxy. In \emph{(a)}, only galaxies with $M_{\star}>10^{10}M_{\sun}$ are used; \emph{(b)} uses lower mass galaxies with $M_{\star}>10^{8.5}M_{\odot}$. Since the segregation results for the DM-halo masses of the galaxies are stronger when low mass galaxies are included, we wanted to determine if the lower mass cut would affect the stellar mass results in the same way. We compare our results to observational results from \citet{Roberts15}, shown in black; while their sample is complete and unweighted in \emph{(a)}, it is $V_{\text{max}}$ weighted in \emph{(b)} to correct for incompleteness. The mass limits on the host haloes are identical between our results and the observational results. The errors shown are standard errors on the mean. Note that we use $r_{vir}$ instead of $R_{200}$ and our positions are 3-D while those in \citet{Roberts15} are projected. } \label{fig:mstarSeg}
\end{figure*}

\subsection{Comparing $M_{\text{z=0}}$ and $M_{\text{peak}}$}

We first investigate the mass segregation of the galaxies in terms of both present-day mass, $M_{\text{z=0}}$, as well as peak mass, $M_{\text{peak}}$. In Fig.~\ref{fig:massSegMbins} (a) \& (b), we show the average galaxy mass as a function of halo-centric radius.  The different colours represent bins of $\log{M}$ where the mass is defined as $M_{\text{z=0}}$ in (a) and $M_{\text{peak}}$ in (b). The upper limit for the bins is always set to $12.5$, while the lower limit is gradually altered to show the effect of excluding low mass galaxies.

The results in Fig.~\ref{fig:massSegMbins}(a) were reported in our previous study \citep{Joshi16} and reproduced here for comparison. Note that the sample used here is slightly different than the one in our previous study due to the additional processing carried out by \textsc{Consistent Trees} (see Sec. \ref{sec:code}). However, these differences are negligible for our results. We had previously found a statistically significant but mild negative trend in $M_{\text{z=0}}$ in the inner $0.5\,r_{\text{vir}}$, i.e. average $M_{\text{z=0}}$ decreasing with halo-centric radius. The trends were weaker when low mass galaxies were excluded. Both these results are reproduced here.

The results in Fig. \ref{fig:massSegMbins}(b), which show segregation in $M_{\text{peak}}$, are consistent with those in Fig. \ref{fig:massSegMbins}(a), showing that the mass segregation trends do not depend on the mass definition we use. This is mainly because, for each point on the plots, we average over the same absolute values of mass. This implies that within each radial bin, there are no significant differences in the distribution of $M_{\text{z=0}}$ and $M_{\text{peak}}$ for the galaxies. This does not mean that the two samples are equivalent however. To understand these results further, we examine the trends in the fractional mass retained since peak to present-day, $M_{\text{z=0}}/M_{\text{peak}}$, in Fig.~\ref{fig:massSegMbins} (c) \& (d), where the samples used are the same as in Fig. \ref{fig:massSegMbins} (a) \& (b) respectively. Fig.~\ref{fig:massSegMbins} (c) \& (d) show that the trend in the ratio of peak mass to present-day mass is much stronger than the trend in absolute mass. There is a strong trend in both Fig.~\ref{fig:massSegMbins} (c) \& (d) with $M_{\text{z=0}}/M_{\text{peak}}$ rising with radius out to around $\sim r_{\text{central}}$, beyond which the trends are shallower. Additionally, the $M_{\text{z=0}}$-selected galaxies consistently retain more of their peak mass compared to the $M_{\text{peak}}$-selected galaxies at all radii. These results show that the $M_{\text{peak}}$-selected sample of galaxies has undergone more mass loss compared to the $M_{\text{z=0}}$-selected sample.

It should be noted that the weak mass segregation trends seen here could result naturally if the halo finder were systematically unable to detect smaller galaxies near the centres of their host haloes. Previous work has shown that the radial distribution of subhaloes in dark matter simulations is less concentrated than that of observed satellite galaxies \citep[e.g. see][etc.]{Gao04,Springel08,Budzynski12}. We explored this issue in more detail in our previous work \citep{Joshi16} and found that \textsc{Rockstar} was better able to recover low mass subhaloes near the centres of their host haloes compared to another popular halo finder, \textsc{Ahf}, which only uses spatial information \citep{Knollmann09}. Additionally, \textsc{Consistent Trees} was designed to follow subhaloes across timesteps more consistently to ensure that we do not lose subhaloes when they pass close to their host halo's centre \citep{Behroozi13b}. We examined the radial distributions of the number density of our galaxy samples (not shown here) and while we find some flattening within $\sim0.5\,r_{\text{vir}}$, the flattening is independent of galaxy mass. This lack of a \emph{differential} flattening between the low-mass and high-mass galaxies implies that the average mass in each radial bin is not significantly affected. This radial flattening independent of subhalo mass was also seen by \citet{Springel08}. These factors would suggest that our mass segregation trends are not numerical artifacts.

One additional result to note from Fig.~\ref{fig:massSegMbins} (c) \& (d) is that the mass loss within each radial bin is nearly independent of the mass of the galaxies ($M_{\text{z=0}}$ in (c) and $M_{\text{peak}}$ in (d)), which implies that any mass segregation trends are not due to the preferential loss of smaller galaxies in the inner regions of the host haloes. Instead, this result confirms that the mass segregation trends we see are due to larger galaxies having formed closer to the centres of their final host haloes.

The galaxies we consider here have spent a significant amount of time in dense environments and therefore, could have lost mass due to tidal stripping, which removes the more extended dark matter from a galaxy more efficiently than its more bound stellar content. It follows that when comparing with observational results, $M_{\text{peak}}$ is the more physically motivated choice as $M_{\text{peak}}$ is much more correlated with present day stellar mass than $M_{\text{z=0}}$, as has been shown in previous studies. For example, using N-body simulations of dwarf spheroidals in clusters, \citet{Penarrubia08} showed that nearly $\sim90\%$ of the galaxies' dark matter had to be tidally stripped before any of the stellar content was stripped. More recently, \citet{Smith16} used hydrodynamical simulations of cluster galaxies with a wide range of stellar masses to show that when tidal stripping had removed $\sim80\%$ of dark matter from the galaxies, they had only lost $\sim10\%$ of their stellar content. Therefore, for the remainder of this study, we focus on the sample of $M_{\text{peak}}$-selected galaxies.

\subsection{Comparing to observational results} \label{sec:obs}

In Fig. \ref{fig:mstarSeg}, we show the expected mass segregation in terms of stellar mass, $M_{\star}$. We assign $M_{\star}$ to the galaxies using two different prescriptions for the stellar-mass-to-halo-mass relations, $f_{\star} = M_{\star}/M_{\text{halo}}$, to convert $M_{\text{peak}}$ to a stellar mass. The first is from \citet{Hudson15} (hereafter H15):
\begin{equation}
f_{\star}(M_{\text{halo}}) = 2f_{\text{1}}\left[ \left(\frac{M_{\text{halo}}}{M_{\text{1}}}\right)^{-\beta} + \left(\frac{M_{\text{halo}}}{M_{\text{1}}}\right)^{\gamma} \right]^{-1}
\end{equation}
where we use $M_{\text{halo}} \equiv M_{\text{peak}}$. Although their best-fitting model suggests a mild evolution in $f_{\star}$ with redshift, the range of redshifts considered in H15 is $0.2-0.8$, whereas the analogue sample in our study covers a range in $z_{\text{peak}}$ of $0-5.5$. The parameters for the best-fitting model would result in a value of $f_{\star}$ that exceeds the cosmic baryon fraction by $z_{\text{peak}}\sim4.5$. We therefore use the parameters for their `no-evolution' model for all galaxies: $f_{\text{1}} = 0.04$, $\log{M_{\text{1}}} = 12.38$, $\beta = 0.69$ and $\gamma=0.8$, and do not include scatter in the relation.

The second relation we use is from \citet{Behroozi13c} (hereafter B13):
\begin{align}
\log{(M_{\star}(M_{\text{h}}))} = \log{(\epsilon M_{1})} + f\left( \log{\left( \frac{M_{\text{h}}}{M_{1}} \right)} \right) - f(0) \\
f(x) = -\log{(10^{\alpha x}+1)} + \delta\frac{(\log{(1+\exp{(x)})})^{\gamma}}{1+\exp{10^{-x}}}
\end{align}
where the parameters $M_{1}$, $\alpha$, $\gamma$, $\delta$ and $\epsilon$ are functions of redshift. We use the best fit parameters from \citet{Behroozi13c} at $z_{\text{peak}}$ and $M_{\text{h}} \equiv M_{\text{peak}}$. Again, we do not introduce any scatter in the relation.

Fig.~\ref{fig:mstarSeg} shows average profiles for $M_{\star}$ as a function of halo-centric radius for two lower mass limits and for the two different stellar-mass-to-halo mass relations. In order to compare with observational results, we also show results from \citet{Roberts15} in black. The observational results in Fig.~\ref{fig:mstarSeg}(b) have been $V_{\text{max}}$ weighted to account for incompleteness at such low stellar masses. There is a systematic offset between our results and the observations; however, the trends differ by at most $0.3$ dex using the H15 relation, $0.15$ using the B13 relation, and have similar slopes. The stellar mass functions for our data follow a single power law, similar to their (dark-matter) halo mass functions as shown in \citet{Joshi16} (figure 2),  whereas those for the observational data have characteristic Schecter function profiles, which could be one of the factors responsible for the differences between our results and the observations. It should be noted that amongst recent studies that find evidence for mass segregation, the trends detected are quite weak, which is consistent with our findings.

\section{The role of preprocessing}	\label{sec:preproc}

\begin{figure}
\includegraphics[width=\linewidth]{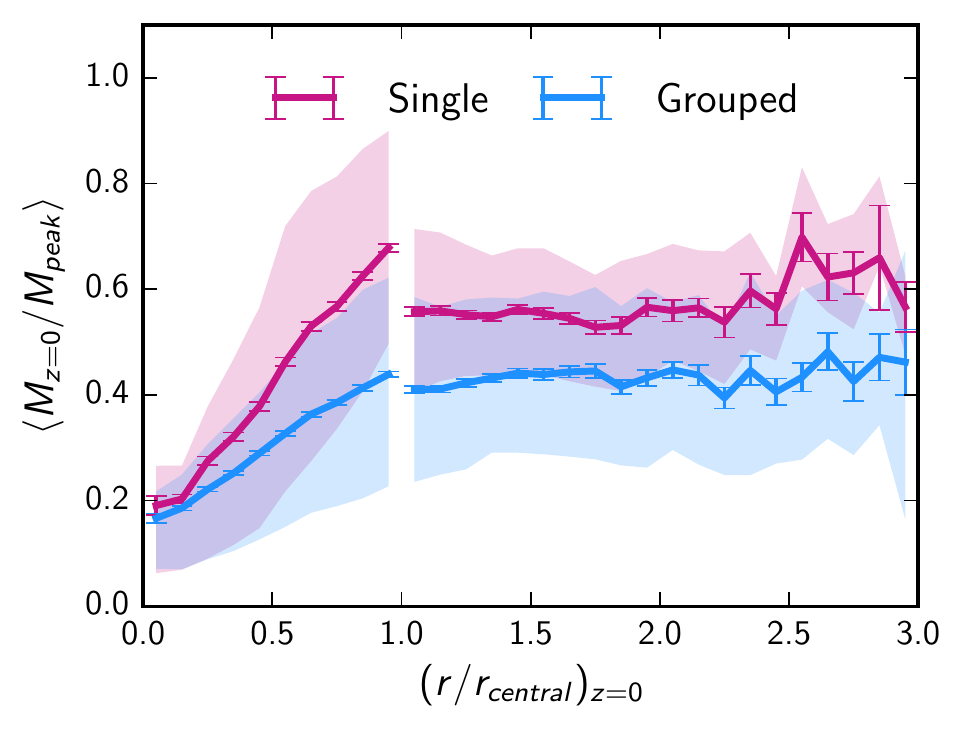}
\caption{Average mass of the galaxies retained since peak to present-day as a function of present-day halo-centric radius, as in Fig. \ref{fig:massSegMbins}(d), now separated into single (pink) and grouped (blue) samples. The shaded areas show the $25^{th}-75^{th}$ percentile range of the data in each radial bin whereas the errorbars show the standard error on the mean. The gap at $r_{\text{central}}$ separates the infall and backsplash populations.} \label{fig:preproc}
\end{figure}

\begin{figure}
\includegraphics[width=\linewidth]{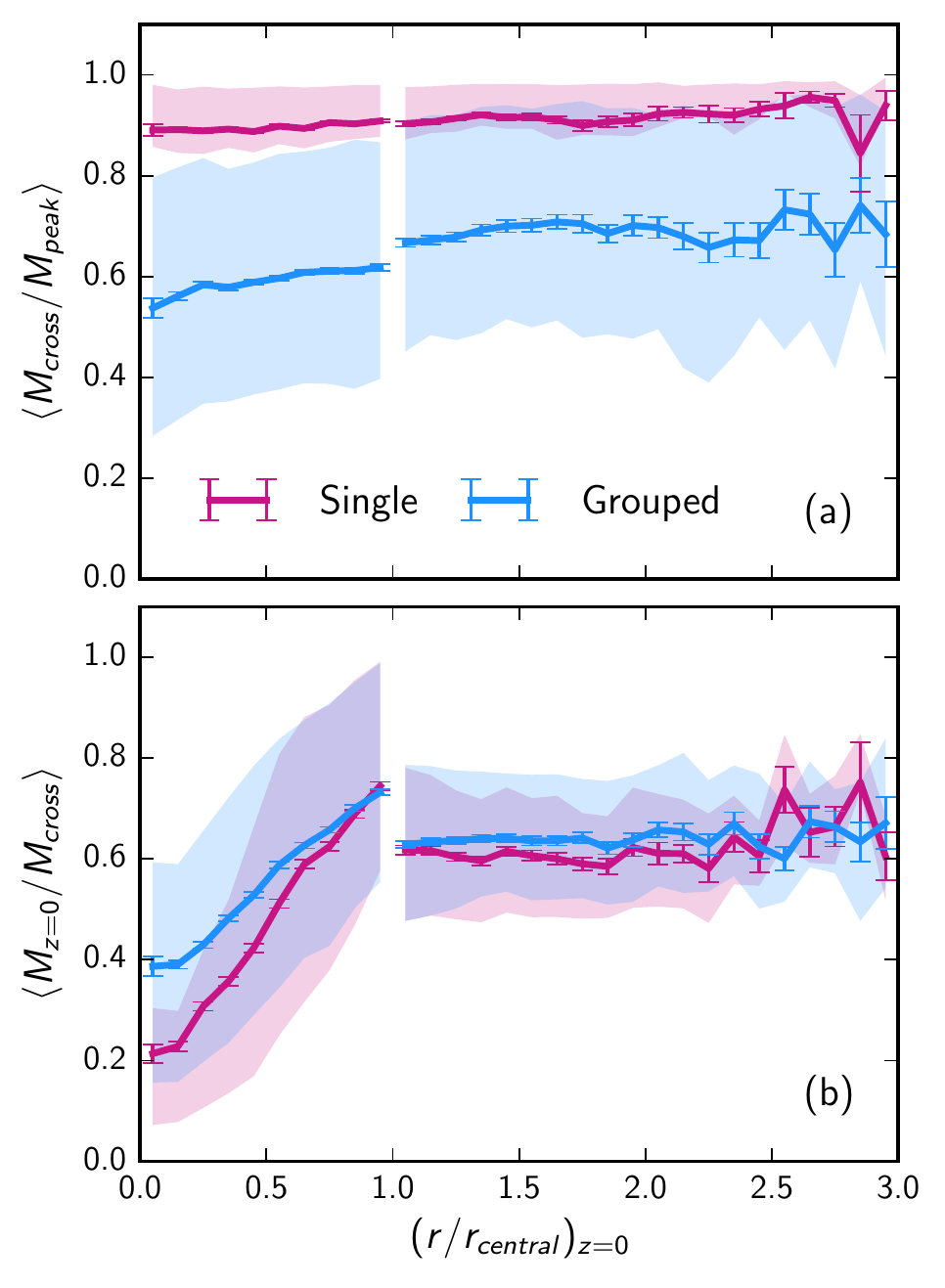}
\caption{Average mass of the galaxies retained from peak to crossing \emph{(a)} and from crossing to present-day \emph{(b)} as a function of present-day halo-centric radius. The single sample is shown in pink, the grouped sample in blue. The shaded areas show the $25^{th}-75^{th}$ percentile range of the data in each radial bin whereas the errorbars show the standard error on the mean. The gap at $r_{\text{central}}$ separates the infall and backsplash populations. \emph{(a)} shows clear evidence for preprocessing where the grouped galaxies have consistently undergone more mass loss than the single galaxies \emph{before} they crossed within a virial radius of the central halo.} \label{fig:preprocStep}
\end{figure}

\begin{figure}
\includegraphics[width=\linewidth]{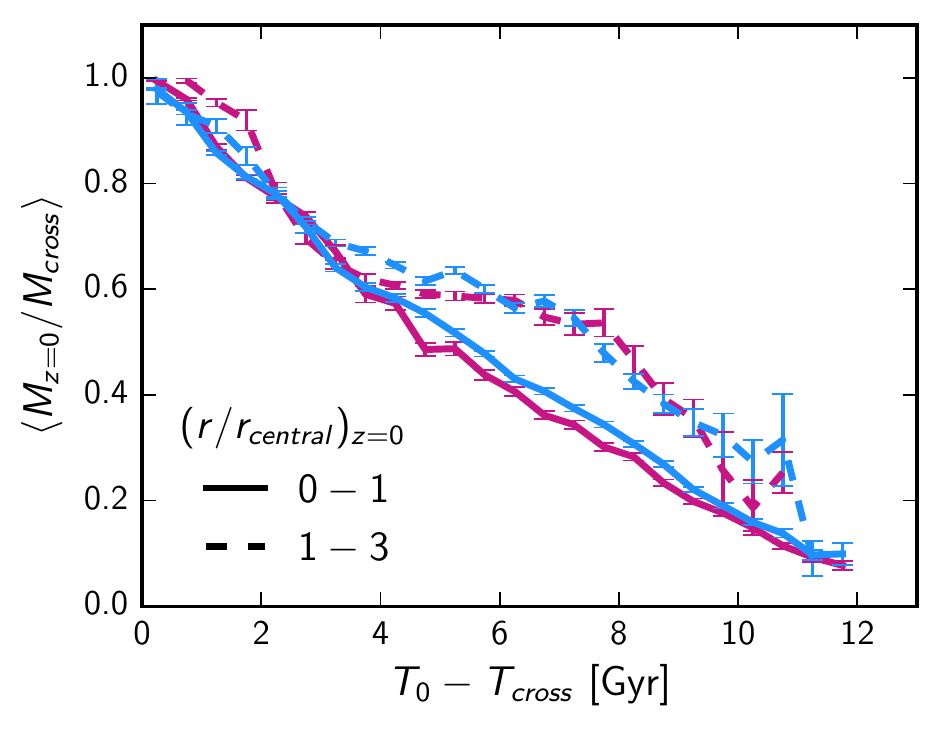}
\caption{Average mass of the galaxies retained from crossing to present-day as a function of time since crossing ($T_{0}$ and $T_{cross}$ are the age of the Universe at $z=0$ and $z_{\text{cross}}$ respectively). The single sample is shown in pink, the grouped sample in blue, as in Fig. \ref{fig:preproc}. The different linestyles are bins of present-day radial position of the galaxies in their central haloes such that the backsplash galaxies are shown in the dashed lines. No significant difference in seen in the mass lost by the single and grouped galaxies after crossing the virial radius of their central halo.}\label{fig:preprocACbyR}
\end{figure}

\begin{figure}
\includegraphics[width=\linewidth]{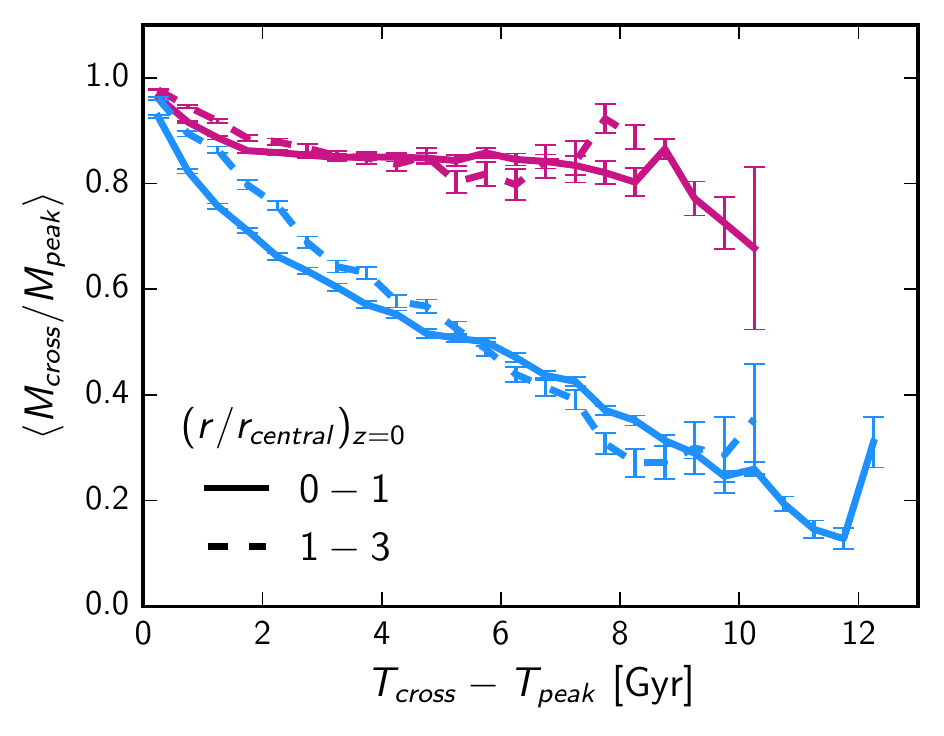}
\caption{Average mass of the galaxies retained from peak to crossing as a function of time between peak and crossing events ($T_{\text{peak}}$ and $T_{\text{cross}}$ are the age of the Universe at $z_{\text{peak}}$ and $z_{\text{cross}}$ respectively). The single sample is shown in pink, the grouped sample in blue, as in Fig. \ref{fig:preproc}. The different linestyles are bins of present-day radial position of the galaxies in their central haloes such that the backsplash galaxies are shown in the dashed lines. The grouped sample has lost significantly more mass before crossing as compared to the single sample. The mass lost is strongly correlated with the time interval between $z_{\text{peak}}$ and $z_{\text{cross}}$ with the trend being much steeper for the grouped sample and nearly flat for the single sample.} \label{fig:preprocBCbyR}
\end{figure}

The results of the previous section show that galaxies undergo significant mass loss from their peak mass under the influence of a group/cluster halo. In this section, we explore whether the galaxies experience some of this mass loss in smaller haloes before they become part of their final group/cluster. 

To do so, we first separate the galaxy sample into galaxies that were part of a smaller group prior to infall and those that were not. We define these samples based on whether or not the analogue was distinct (i.e. identified as the top of the subhalo hierarchy by the halo finder) before the first time it crosses within a virial radius. We first remove those galaxies that have never crossed within $r_{\text{vir}}$ of their central halo, which affects a little less than half our sample, reducing the sample size to $45105$. Galaxies found beyond a virial radius at $z=0$ are necessarily backsplash galaxies. Such galaxies constitute $26\%$ of this reduced sample ($87\%$ of the backsplash galaxies lie between $(1-2)\,r_{\text{central}}$ while the remaining $13\%$ lie beyond $2\,r_{\text{central}}$), and represent a different population compared to those found within the virial radius. Hence, for any further analysis, we treat the two separately. Also note that not all galaxies found within a virial radius are on their first passage within the central halo, nor are all of them on an infall trajectory (i.e. some are on their way out of the central halo); but for convenience, we will simple refer to this subsample as the `infall' population. We separate the galaxies into those that were distinct at \emph{all} timesteps before the first time they crossed the virial radius of their central halo and those that were not (we hereafter refer to this event as `crossing'). The former population is designated as `single', the latter as `grouped'. Most of the galaxies reach their peak mass before or at the time of crossing; however a small fraction of them, $\sim15\%$ of our total sample ($24\%$ of the single sample, $10\%$ grouped sample), do so after crossing. For clarity, we remove these galaxies from the sample, although including them does not qualitatively change our subsequent results. We are then left with $38287$ galaxies in total, $12666$ in the single category, $25621$ in the grouped category. 

In Fig. \ref{fig:preproc}, we first examine the average total mass retained from peak to present-day, $M_{\text{z=0}}/M_{\text{peak}}$, as a function of present-day halo-centric radius, as in Fig. \ref{fig:massSegMbins}(d), now separating the sample into single and grouped populations. The shaded areas show the $25^{th}-75^{th}$ percentile range of the data in each radial bin, while the errorbars are standard errors on the mean. The gap at $r_{\text{central}}$ separates the infall and backsplash galaxies. The figure shows that at nearly all radii, the grouped galaxies have lost more mass since their peak compared to the single galaxies, which suggests that the grouped galaxies were preprocessed. For the backsplash galaxies, the differences between the single and grouped galaxies are nearly constant at all radii with the grouped galaxies losing $\sim15-20\%$ more mass than the single galaxies. However, for the infall populations, the differences between the grouped and single galaxies are negligible at the halo centre and increase with increasing radius. The negligible difference at the halo centre appears to be at odds with the significant $\sim15-20\%$ difference for the backsplash galaxies, since the backsplash galaxies would have passed near the halo centre at an earlier time and should now display the same amount of mass loss between the single and grouped populations.

To understand the exact sequence of mass loss and gain, we separate the mass loss since peak into two stages -- before and after crossing. We show the average mass retained from peak to crossing, $M_{\text{cross}}/M_{\text{peak}}$, in Fig. \ref{fig:preprocStep}(a), and crossing to present-day, $M_{\text{z=0}}/M_{\text{cross}}$, in Fig. \ref{fig:preprocStep}(b), as a function of halo-centric radius. Note that while $M_{\text{cross}}/M_{\text{peak}}$ is less than one by definition for all galaxies, $M_{\text{z=0}}/M_{\text{cross}}$ can be greater than one. Fig.~\ref{fig:preprocStep}(b) shows that, within a virial radius, the closer the galaxies are to the central halo, the more mass they appear to have lost. The single galaxies lose \emph{more} mass than the grouped galaxies. The differences are largest at small radii and decrease with increasing radius. This can be explained by the fact that grouped galaxies are in a denser potential and therefore somewhat shielded from the effects of the final host halo. Additionally, if they have already undergone some mass loss in smaller haloes, they are likely to be compact, dense objects, less prone to mass loss. The single galaxies on the other hand are likely to have much more loosely bound material in their outskirts which is more easily stripped by the host halo. The results beyond a virial radius are consistent with having no radial trend and there is no significant difference between the single and grouped sample. This may suggest that after passing through pericentre, the galaxies do not retain any information regarding whether they were single or grouped, or regarding their orbital properties, although more analysis is needed to verify this. 

However, \emph{before} crossing a virial radius, there is a clear difference in the amount of mass lost since peak, as shown in Fig.~\ref{fig:preproc}(a). The grouped sample has consistently lost more mass than the single sample; for the single sample, $\left\langle M_{\text{cross}}/M_{\text{peak}} \right\rangle \sim0.88$ whereas for the grouped sample, $\left\langle M_{\text{cross}}/M_{\text{peak}} \right\rangle \sim 0.55-0.65$. This is clear evidence of galaxies being preprocessed in smaller groups before they are accreted onto their present-day host haloes. Figs. \ref{fig:preprocStep} (a) \& (b) therefore suggest a scenario in which, compared to single galaxies, grouped galaxies lose significantly more mass in smaller haloes before becoming part of their final host haloes through preprocessing. However, once both sets of galaxies cross within a virial radius of the host halo, the single galaxies lose more mass compared to the grouped galaxies, essentially `catching up' with the total amount of mass loss of the grouped galaxies. This sequence of events also explains why the backsplash galaxies have nearly the same amount of mass loss after crossing whereas they have significant differences in the total amount of mass loss since peak, as seen in Fig. \ref{fig:preproc}.

To investigate if the results in Figs. \ref{fig:preproc} \& \ref{fig:preprocStep} are driven by other properties of the galaxies, we examined these radial trends in bins of present day galaxy mass $M_{\text{z=0}}$, the mass of the central halo $M_{\text{central}}$, the peak redshift $z_{\text{peak}}$, and the redshift of crossing $z_{\text{cross}}$. We find no dependence of the average mass loss (before or after accretion) on $M_{\text{z=0}}$, and a mild dependence on $M_{\text{central}}$ where galaxies in more massive clusters had lost slightly more mass before crossing, and slightly less after crossing, as compared to galaxies in lower-mass groups. These results are expected since galaxies found in more massive clusters at $z=0$ are more likely to be preprocessed as the cluster is likely to have accreted several groups of galaxies. However, the results showed a strong dependence on $z_{\text{peak}}$ and $z_{\text{cross}}$, implying that the main factor driving these results is the amount of time the galaxies have spent within a given environment. Hence, we examine this in further detail in Figs. \ref{fig:preprocACbyR} and \ref{fig:preprocBCbyR}.

In Fig. \ref{fig:preprocACbyR}, we show the average mass retained after crossing as a function of time since crossing. The infall sample is shown by the solid lines, the backsplash sample by the dashed lines. The figure confirms that there is no significant difference between the single and grouped galaxies once they have crossed within a virial radius. The infall population displays a strong, nearly monotonic trend with time since crossing; these galaxies appear to steadily lose mass within the central halo. The bump seen in the trends for the backsplash galaxies at longer times since crossing is likely due to the fact that these galaxies have spent a significant amount of time outside a virial radius of the central halo and therefore experienced lower amounts of mass loss. Galaxies who first crossed a virial radius $\lesssim 3$Gyr ago have probably only recently crossed back out.

Fig. \ref{fig:preprocBCbyR} shows the average mass retained from peak to crossing as a function of the time interval between peak and crossing. As in Fig. \ref{fig:preprocACbyR}, the infall population is shown by the solid lines, the backsplash population by the dashed lines. In this figure, there is a significant separation between the single and grouped samples as in Fig. \ref{fig:preprocStep}(a), which is clear evidence for preprocessing. The amount of mass lost is strongly correlated with the time interval between peak and crossing for the grouped sample, whereas for the single sample, the trend is nearly flat with the galaxies losing $\sim10\%$ of their peak mass regardless of the time interval. This mass loss may partially be due to how well the halo finder can recover the masses of these galaxies at each timestep. Note that by definition, any uncertainty in galaxy mass will be seen as a mass loss when compared to peak mass. \citet{Knebe11} investigated the performance of several halo finders by running them on a dark-matter system consisting of a halo, subhalo and subsubhalo with known properties and found that \textsc{Rockstar} was able to recover their masses to within $\sim10\%$.

\section{Discussion}	\label{sec:disc}

\subsection{Mass segregation}
In this work, we have examined mass segregation in group/cluster haloes. In our previous study using dark-matter simulations \citep{Joshi16}, we showed that the mass segregation of galaxies was restricted to within $\sim0.5\,r_{\text{vir}}$ where galaxies near the centres of groups and clusters were on average more massive than those at large radii. Here we apply a more physically motivated mass definition for the galaxies, $M_{\text{peak}}$, and re-examine these radial trends. Using $M_{\text{peak}}$ also allows us to directly compared to observational results that use stellar mass. We compare the radial trends using both mass definitions and find that they both produce the same results within the mass range of interest. However, the two mass definitions select significantly different populations of galaxies -- the $M_{\text{z=0}}$-selected sample consisted of high peak-mass galaxies that had undergone a lower amount of mass loss, the $M_{\text{peak}}$-selected sample consisted of intermediate peak-mass galaxies that had undergone a higher amount of mass loss. This is apparent in Figs.~\ref{fig:massSegMbins} (c) \& (d), where the $M_{\text{z=0}}$-selected sample has values of $M_{\text{z=0}}/M_{\text{peak}}$ that are consistently higher than for the $M_{\text{peak}}$-selected sample at all radii and irrespective of galaxy mass. The results in Figs. \ref{fig:massSegMbins} (a) \& (b) are broadly in agreement with \citet{vanDenBosch16} who found mild segregation in both  $M_{\text{z=0}}$ and  $M_{\text{peak}}$, although note that they explicitly removed any dependence on central halo mass by normalizing these masses by $M_{\text{central}}$. Their sample selection is also based on  $M_{\text{z=0}}/M_{\text{central}}$ as well as mass at accretion, so that our results are not directly comparable.

In our previous study, we concluded that while the results do not rule out the effect of dynamical friction, it was not the most important factor driving the mass segregation trends. The results in Figs.~\ref{fig:massSegMbins} (c) \& (d) show that the galaxies undergo a significant amount of mass loss from peak mass and that there is a strong radial trend in the amount of mass lost. Thus, even if dynamical friction does act on these galaxies, its effects would gradually be diminished with time as the galaxies lose mass due to tidal stripping. We investigated these results in bins of central halo mass and found the same trends as in our previous work. Although not presented here, the mass segregation trends are strongest in low mass groups with $\log{M_{\text{central}}}=[12.5,13]$ and grow weaker with increasing central halo mass. This already suggests that the galaxies are preprocessed in smaller groups before the groups are accreted onto larger clusters, where the infall of groups would destroy any segregation trends that may have existed in the clusters.

\subsection{Preprocessing}	\label{sec:discPreProc}

\begin{figure*}
\includegraphics[width=\textwidth]{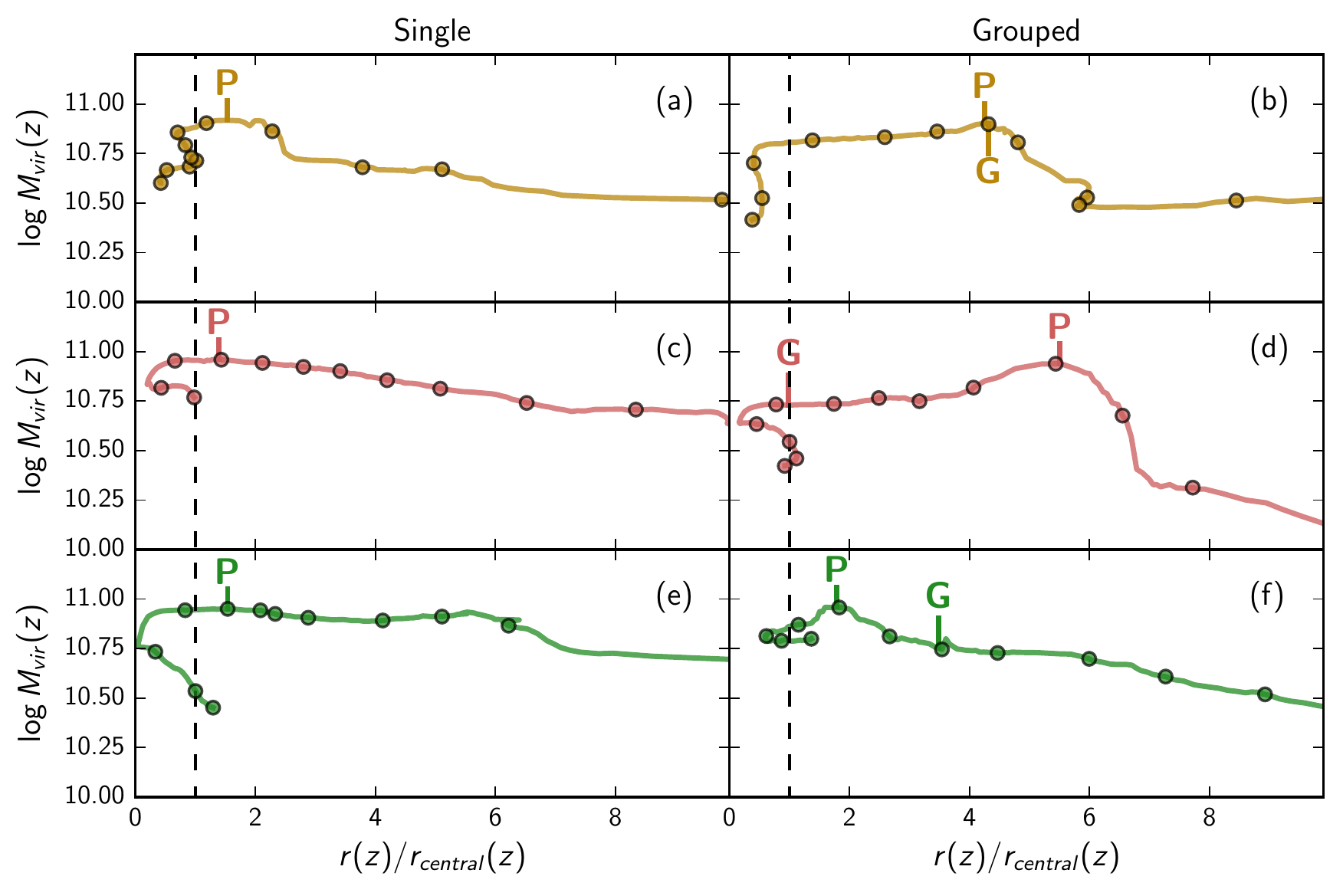}
\caption{Tracks of individual galaxies showing virial mass as a function of distance to the present-day central halo normalized by the central halo's virial radius at that redshift. Three galaxies were chosen randomly within a narrow range in galaxy peak mass and central halo mass ($\log{M_{\text{peak}}}=[10.9-11]$ and $\log{M_{\text{central}}}=[13.5-13.6]$), shown in different colours. The left panel contains galaxies from the single sample, the right from the grouped sample. The black points mark roughly $1\,$Gyr intervals (15 timesteps). The vertical black dashed lines mark $r_{\text{vir}}$ for clarity. We have also marked the peak mass with the letter `P', and for the grouped analogues, the point at which they first become part of a larger halo, marked with the letter `G'.} \label{fig:haloTracks}
\end{figure*}

Fig.~\ref{fig:preprocStep}(a) shows that there is clear evidence for galaxies being preprocessed in smaller groups before being accreted into larger host haloes. We examined the dependence of these results on $M_{\text{peak}}$,  $M_{\text{central}}$,  $z_{\text{peak}}$ and  $z_{\text{cross}}$. We did find a mild correlation with host halo mass for the grouped sample where galaxies in the most massive (present-day) haloes had lost \emph{more} mass before crossing than those in the least massive haloes; the opposite trend is seen for mass lost after crossing. We also found that the crucial variable determining the amount of mass lost by the galaxies is the amount of time they spend in a particular environment. Figs. \ref{fig:preprocACbyR} and \ref{fig:preprocBCbyR} show that the amount of mass lost after crossing is strongly correlated with time since crossing for both the single and grouped sample. In the case of mass loss from peak to crossing however, only the grouped  sample has a similar strong trend with time between peak and crossing; the single galaxies appear to lose $\sim10\%$ of their peak mass regardless of the time interval (provided it is not too short). One factor to keep in mind is that all these results are dependent on the halo finder's ability to consistently detect the galaxies at each time step and reliably determine their masses and sizes, especially in these dense environments. \textsc{Rockstar} is able to detect substructure near host halo centres due to its use of phase-space information, although it can over- or under-predict the subhaloes' masses in such high density environments \citep{Knebe11}. Using \textsc{Consistent Trees} has the advantage of not only consistently following subhaloes, but also to some degree repairing artificial mass fluctuations \citep{Behroozi13b}. Hence, although this is an issue to consider, especially if comparing to results from different halo finders, the results present here are mostly physical and not artifacts produced by the halo finders.

One crucial aspect of this analysis is the radius we chose to define crossing. While a virial radius seems a natural choice, there is evidence that the host halo's influence can begin at a larger distance. In Fig.~\ref{fig:haloTracks}, we show tracks of virial mass as a function of distance from the present-day central halo for individual galaxies. We randomly choose three galaxies each from the single and grouped populations in narrow bins of $\log{M_{\text{peak}}}=[10.9-11]$ and $\log{M_{\text{central}}}=[13.5-13.6]$. This ensures that the chosen galaxies have nearly identical starting points and eventual host environments. Despite this, Fig. \ref{fig:haloTracks} shows that the galaxies can have varied histories and can reach very different present-day masses. Although the single galaxies appear to reach their peak mass close to the virial radius of the host halo, they do not appear to grow significantly in mass from a distance of a few virial radii. The grouped galaxies have more diverse histories. In Fig. \ref{fig:haloTracks}(b), the galaxy appears to peak at the same time it becomes part of a group and then steadily loses mass, which is consistent with the scenario discussed in Section \ref{sec:preproc}, whereby the galaxy continues to grow in mass until it becomes part of a group, after which it steadily loses mass before, as well as after, crossing within a virial radius of its final host halo. In Fig. \ref{fig:haloTracks}(d) however, the galaxy starts losing mass long before it becomes part of a group, suggesting the influence of another halo at distances larger than a virial radius. Fig. \ref{fig:haloTracks}(f) shows a rare scenario where the galaxy continues to grow in mass even after becoming part of a group, although note that such a track is not representative of the sample, rather an interesting outlier. Once the haloes cross a virial radius, regardless of whether they are single or grouped, they steadily lose mass as they spiral towards the host centre. While this analysis is preliminary, it does imply that defining crossing at a virial radius does not adequately capture the effects of the central halo. The degree to which this is important and what the ideal crossing radius definition is will be explored in future work, but it does indicate that any studies involving environmental effects must be careful in defining the radius of accretion as it can significantly alter their results.

\section{Summary}	\label{sec:summ}
We use N-body simulations and define a sample of galaxy analogues to explore mass segregation in groups and clusters as well as the role of preprocessing in determining mass loss trends for these analogues.

\begin{enumerate}
\item Consistent with our results in \citet{Joshi16}, we find weak mass segregation within $0.5\,r_{\text{vir}}$, with average mass decreasing with halo-centric radius. The results are largely independent of the mass definition used -- $M_{\text{z=0}}$ or $M_{\text{peak}}$. We also find a strong radial trend in the amount of mass lost since peak and a significant difference in these trends for the two samples. The $M_{\text{peak}}$-selected galaxies appear to have lost more mass since peak at all radii.
\item We find that grouped galaxies that first cross within a virial radius of their present-day central halo as part of a group lose $\sim 35-45\%$ of their peak mass, compared to single galaxies that cross as distinct haloes, which lose $\sim12\%$. This is clear evidence for preprocessing in smaller haloes before accretion onto the present-day central halo.
\item The fraction of mass lost does not depend on galaxy mass and only weakly depends on the central halo mass. However, we find a strong correlation between the degree of mass loss and the amount of time the galaxies spend in a specific environment.
\end{enumerate}

This study shows that there is clear preprocessing in the mass lost by dark matter haloes. We look forward to including baryonic physics in our simulations which could have additional preprocessing effects. Our future work will focus on following the galaxies along their halo tracks and determining if these results extend beyond a virial radius of the central halo, as well as studying how baryonic processes affect our current results.

\section*{Acknowledgements}
We thank the anonymous referee for their comments and insights which were very useful in improving the manuscript. We thank the National Science and Engineering Research Council of Canada for their funding. Computations were performed on the \emph{gpc} supercomputer at the SciNet HPC Consortium \citep{Loken10}. SciNet is funded by: the Canada Foundation for Innovation under the auspices of Compute Canada; the Government of Ontario; Ontario Research Fund - Research Excellence; and the University of Toronto. This work was made possible by the facilities of the Shared Hierarchical Academic Research Computing Network (SHARCNET:www.sharcnet.ca) and Compute Canada.

\bibliographystyle{mnras}
\bibliography{paper2}


\bsp	
\label{lastpage}
\end{document}